\documentclass[
    ,final            
  ]
  {aipproc}
\usepackage{epsfig}

\layoutstyle{6x9}

\begin{document}

\title{FINITE AXISYMMETRIC CHARGED DUST DISKS SOURCES FOR CONFORMASTATIC
SPACETIMES}


\author{Guillermo A. Gonz\'{a}lez}{
  address={Escuela de F\'isica -- Universidad Industrial de Santander \\
A. A. 678, Bucaramanga, Colombia},
  altaddress={Departamento de F\'isica Te\'orica -- Universidad del Pa\'is Vasco \\
48080, Bilbao, Espa\~na} 
}

\author{Antonio C. Guti\'errez-Pi\~{n}eres}{
  address={Escuela de F\'isica -- Universidad Industrial de Santander \\
A. A. 678, Bucaramanga, Colombia}
}

\author{Paolo A. Ospina}{
  address={Escuela de F\'isica -- Universidad Industrial de Santander \\
A. A. 678, Bucaramanga, Colombia}
}

\begin{abstract}
An infinite family of finite axisymmetric charged dust disks is presented. The
disks are obtained by solving the Einstein-Maxwell equations for conformastatic
spacetimes by assuming a functional dependency between the time-like component
of the electromagnetic potential and the metric potential in terms of a solution
of the Laplace equation. We give solutions to the Einstein-Maxwell equations
with disk sources of finite extension in which the charge density is
proportional to the energy surface density. We apply the well-know ``inverse''
approach to the gravitational potential representing finite thin disks given by
Gonzalez and Reina to generate conformastatic charged dust thin discs. Exact
examples of conformastatic metrics with disk sources are worked out in full.
\end{abstract}

\maketitle


The vacuum Einstein-Maxwell equations
\begin{equation}
G_{ab} = 8 \pi \ T_{ab},  \qquad {F^{ab}}_{;b} = 0, \label{eq:eme}
\end{equation}
with the line element for a conformastatic spacetime \cite{SYN}
\begin{equation}
ds^2 = - \ e^{2\lambda} dt^2  +  e^{- 2\lambda} (r^2 d\varphi^2 + dr^2 + dz^2),
\label{eq:metCC} 
\end{equation} 
where $\lambda$ does not depend on $t$, and if we take $A_a = (- \phi, 0, 0, 0)$,
where $\phi$ also is independent of $t$, reduce to
\begin{eqnarray}
\nabla^2 \lambda &=& e^{- 2 \lambda} \nabla \phi \cdot \nabla \phi,
\label{eq:eme2} \\
\nabla^2 \phi &=& 2 \ \nabla \lambda \cdot \nabla \phi, \label{eq:eme3}\\
\lambda_{,i} \lambda_{,j} &=& e^{- 2 \lambda} \phi_{,i} \ \phi_{,j},
\label{eq:eme1} 
\end{eqnarray}
where $i,j = 1,2,3$ and $\nabla$ is the usual differential operator in
cylindrical coordinates. In order to solve the above system of equations, first
it is assumed that $\phi = \phi(\lambda)$, so that \cite{GGO}
\begin{equation}
\qquad \phi = \pm e^{\lambda} + k_1, \label{eq:philam} 
\end{equation}
 and the equations system (\ref{eq:eme2})-(\ref{eq:eme3}) reduces to
\begin{equation}
\nabla^{2} \lambda = \nabla \lambda \cdot \nabla \lambda. \label{eq:eclam}
\end{equation}
Then, we assume $\lambda = \lambda ( U )$, where $U$ is an auxiliary function
that is taken as a solution of the Laplace equation. Therefore, 
\begin{equation}
e^{\lambda} = \frac{k_{3}}{U + k_{2}}. \label{lambdau} 
\end{equation}
In order to have an appropriated behavior at infinity, we take $k_2 = k_3$
and $k_1 = \mp 1$, so that
\begin{equation}
\phi = \pm \left[ \frac{k}{U + k} - 1 \right], \label{eq:phiu}
\end{equation}
where $k_2 = k_3 = k$. 

Solutions that correspond to finite thin disks can be obtained by introducing
the oblate spheroidal coordinates
\begin{equation} 
r^{2} = a^{2}(1 + \xi^{2})(1 - \eta^{2}), \qquad
z = a \xi \eta, \label{eq:ciloblatas2}
\end{equation}
where $0 \leq \xi < \infty$ and $-1 \leq \eta < 1$. The disk has coordinates
$\xi = 0$, $0\leq\eta^2<1$ and, on crossing the disk, $\eta$ changes sign but
does not change in absolute value. The singular behavior of the coordinate
$\eta$ implies that a polynomial in even powers of $\eta$ is a continuous
function everywhere but has a discontinuous $\eta$-derivative at the disk.
Accordingly, 
\begin{equation}
U(\xi,\eta) = - \sum_{n=0}^{\infty} C_{2n} q_{2n}(\xi) P_{2n}(\eta),
\label{eq:gensol}
\end{equation}
where the $P_{2n}(\eta)$ are the legendre polynomials of order $2n$ and
$q_{2n}(\xi) = i^{2n+1} Q_{2n}(i\xi)$, with $Q_{2n}(i\xi)$ the Legendre function
of second kind of imaginary argument \cite{BAT}. By using the distributional approach \cite{PH, LICH, TAUB}, the Surface
Energy-Momentum Tensor and the Surface Current Density of the disk can be
written as
\begin{equation}
S^{ab} = \epsilon V^a V^b,  \qquad j^a = \sigma V^a,
\end{equation}
where $V^a = e^{-\lambda} (1, 0, 0, 0 )$ is the velocity vector. The energy
density and the charge density are given, respectively, by
\begin{equation}
\epsilon =  \frac{e^\lambda \lambda_{,z}}{2 \pi} , \qquad \sigma =  -
\frac{\phi_{,z}}{2 \pi} \label{eq:sigma} .
\end{equation}
Now, by using (\ref{eq:philam}), is easy to see that
\begin{equation}
\sigma = \mp \ \epsilon  \label{eq:sigma*}
\end{equation}
so that the charge density of the disks is equal, up to a sign, to their mass
density. 

From equation (\ref{lambdau}), we have that 
\begin{equation}
\epsilon = - \frac{k \Sigma}{(U + k)^2} \ , \qquad \Sigma = \frac{U_{,z}}{2 \pi}
\end{equation}
$\Sigma$ is the Newtonian mass density of a disklike source. We shall consider a
family of well behaved Newtonian thin disks of finite radius, the generalized
Kalnajs disks \cite{GR}, whose Newtonian densities are given by
\begin{equation}
\Sigma_{m} (r) = \frac{(2m+1)M}{2\pi a^{2}} \left[ 1 - \frac{r^{2}}{a^{2}}
\right]^{m - \frac{1}{2}} .
\end{equation}
$M$ and $a$ are the total mass and the radius of the disk and we must take $m
\geq 1$. The constants $C_{2n}$ are defined through the relation
\begin{equation}
C_{2n} =  \frac{K_{2n}}{(2n+1) q_{2n+1}(0)},
\end{equation}
where
\begin{equation}
K_{2n} = \frac{M}{2a} \left[ \frac{ \pi^{1/2} \ (4n+1) \ (2m+1)!}{2^{2m} (m -
n)! \Gamma(m + n + \frac{3}{2})} \right] 
\end{equation}
for $n \leq m$, and $C_{2n} = 0$ for $n > m$. The dimensionless quantity ${\tilde \epsilon}_m ({\tilde r}) = \pi
a \epsilon_m ({\tilde r})$ for the first four members of the family are
\begin{eqnarray}
{\tilde \epsilon}_1 &=& - \frac{3 {\tilde k} \sqrt{1 - {\tilde r}^2}}{2 [{\tilde
k} + \frac{3 \pi}{8} ({\tilde r}^2 - 2)]^2}, \\
	&&	\nonumber	\\
{\tilde \epsilon}_2 &=& - \frac{5 {\tilde k} (1 - {\tilde r}^2)^{3/2}}{2
[{\tilde k} - \frac{15 \pi}{128} (3 {\tilde r}^4 - 8 {\tilde r}^2 + 8)]^2}, \\
	&&	\nonumber	\\
{\tilde \epsilon}_3 &=& - \frac{7 {\tilde k} (1 - {\tilde r}^2)^{5/2}}{2
[{\tilde k} + \frac{35 \pi}{512} (5 {\tilde r}^6 - 18 {\tilde r}^4 + 24 {\tilde
r}^2 - 16)]^2}, \\
	&&	\nonumber	\\
{\tilde \epsilon}_4 &=& - \frac{9 {\tilde k} (1 - {\tilde r}^2)^{7/2}}{2 [
{\tilde k} - \frac{315 \pi}{32768} (35 {\tilde r}^8 - 160 {\tilde r}^6 + 288
{\tilde r}^4 - 256 {\tilde r}^2 + 128)]^2}, \qquad
\end{eqnarray}
where ${\tilde k} = (k a)/M$, ${\tilde r} = r/a$, $0 \leq {\tilde r} \leq 1$.

\begin{figure}
$\begin{array}{cc}
\epsfig{width=2.95in,file=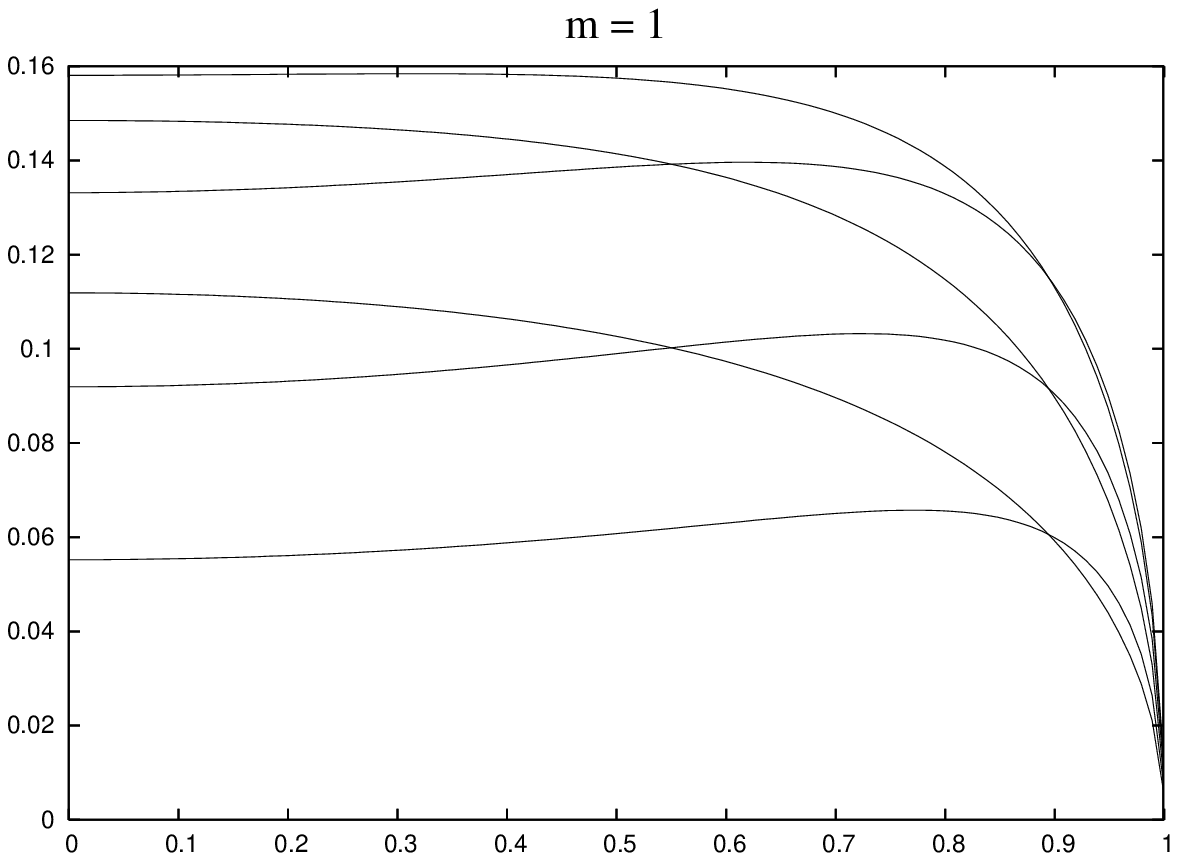} &
\epsfig{width=2.95in,file=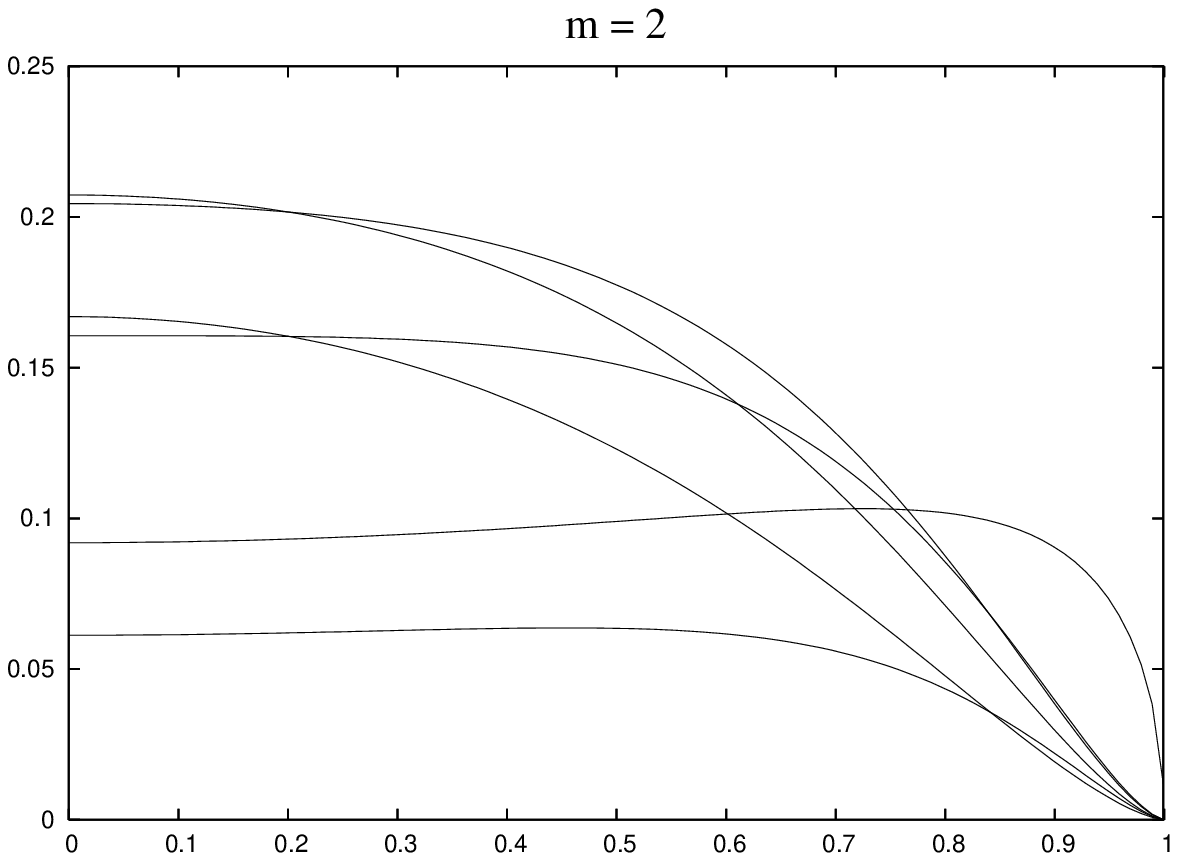} \\
\epsfig{width=2.95in,file=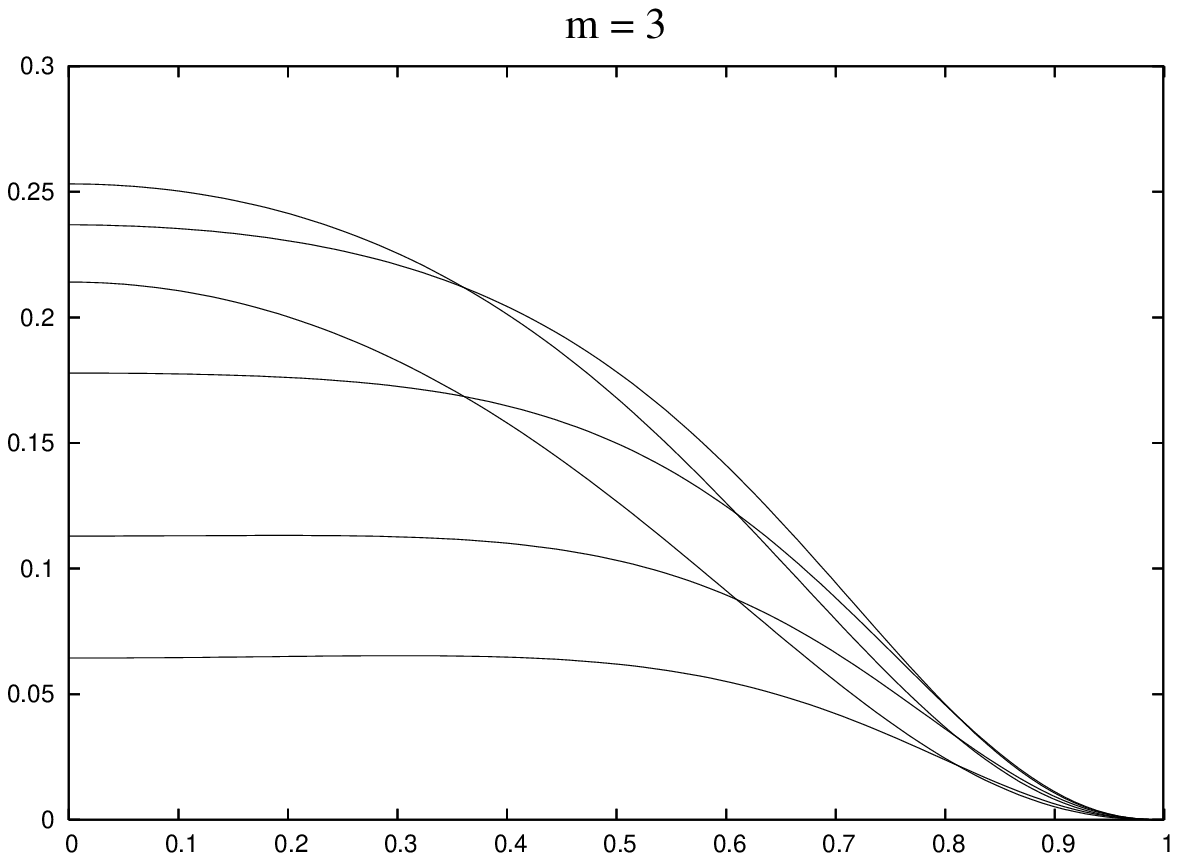} & 
\epsfig{width=2.95in,file=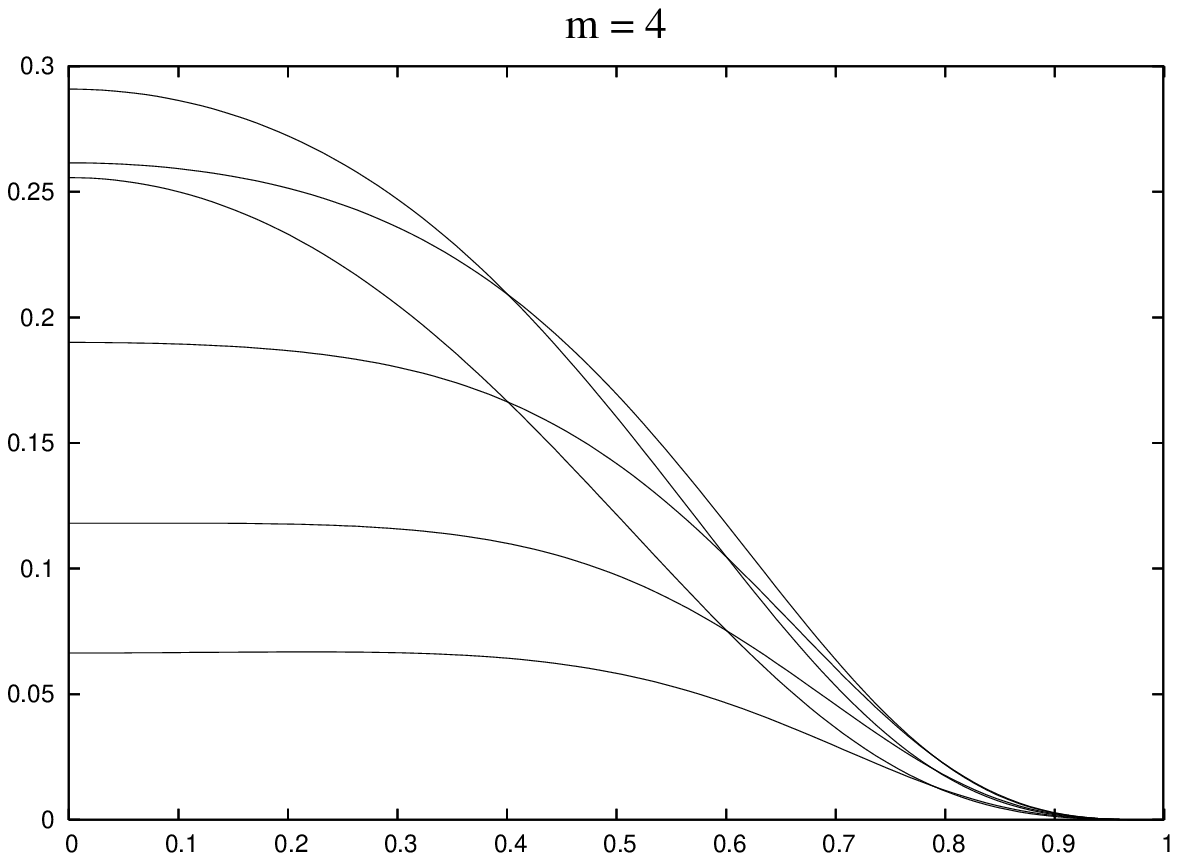} \\
\end{array}$
\caption{${\tilde \epsilon} = a \epsilon$ as a function of ${\tilde r} = r/a$ 
for the disk with $n= 2$ and $\alpha$ = 0.4, 0.7, 1, 1.3 and 1.5.}
\end{figure}

The relativistic thin disks here presented have a charge density that is equal,
up to a sign, to their energy density, and so they are examples of the commonly
named `electrically counterpoised dust' equilibrium configuration. The energy
density of the disks is everywhere positive and well behaved, vanishing at the
edge. Also, as the value of $m$ increases, the energy density is more
concentrated at the center of the disks, having a maximum at $r = 0$ for all the
values of $|{\tilde k}|$. However, for the first two models with $m = 1$ and $m
= 2$, for small values of $|{\tilde k}|$ the energy density presents a maximum
near the edge of the disk whereas that for higher values of $|{\tilde k}|$ the
maximum occurs at the center of the disk. Furthermore, as the energy density of
the disks is everywhere positive and the disks are made of dust, all the models
are in a complete agreement with all the energy conditions, a fact of particular
relevance in the study of relativistic thin disks models.


\begin{theacknowledgments}
A. C.~Guti\'errez-Pi\~{n}eres wants to thank the financial support from
COLCIENCIAS, Colombia.
\end{theacknowledgments}



\bibliographystyle{aipproc}   


\end{document}